\documentclass[5p,times,twocolumn]{elsarticle}

\usepackage{amsmath,amssymb,amsfonts}
\usepackage{graphicx}
\usepackage{booktabs}
\usepackage{multirow}
\usepackage{subcaption}
\usepackage{xcolor}
\usepackage{hyperref}

\journal{Physics Letters B}

\begin{document}

\begin{frontmatter}

\title{Kruglov-entropy cosmology from apparent-horizon thermodynamics: background dynamics and observational constraints}

\author[nu,mu]{Candrasyah Muhammad}
\ead{candramuhammad91@gmail.com}
\author[mu]{Burin Gumjudpai}
\ead{burin.gum@mahidol.ac.th}
\author[mu]{Nandan Roy}
\ead{nandan.roy@mahidol.ac.th (Corresponding Author)}
\author[nu2]{Pornrad Srisawad}
\ead{pornrads@nu.ac.th}

\address[nu]{The Institute for Fundamental Study ``The Tah Poe Academia Institute'', Naresuan University, Phitsanulok 65000, Thailand}
\address[mu]{NAS, Centre for Theoretical Physics \& Natural Philosophy, Mahidol University, Nakhonsawan Campus, Phayuha Khiri, Nakhonsawan 60130, Thailand}
\address[nu2]{Department of Physics, Naresuan University, Phitsanulok 65000, Thailand}

\begin{abstract}
We investigate two late-time cosmological models generated by Kruglov's nonadditive entropy at the apparent horizon. Model I, originally derived by Kruglov, follows the Cai--Kim thermodynamic construction and leads to an implicit modified Friedmann equation, whereas Model II, proposed and derived in this work, is obtained from an entropy-integral prescription and contains an effective vacuum contribution. We formulate both models in terms of the dimensionless Hubble parameter, include radiation, impose exact present-day closure, and evolve the physical branch using analytic derivatives of the implicit Friedmann equations. We constrain the models with DESI DR2 baryon acoustic oscillations and a compressed CMB likelihood, with and without DES-Y5 Type-Ia supernovae. Model I is driven to a markedly lower Hubble constant, $H_0\simeq62.4,\mathrm{km,s^{-1},Mpc^{-1}}$ for the combined data, and remains non-phantom with $w_{\rm DE,0}\simeq-0.814$; it is strongly disfavoured relative to flat $\Lambda$CDM. Model II stays close to $\Lambda$CDM in the standard cosmological parameters and gives $w_{\rm DE,0}\simeq-1.003$, while its two entropy-sector parameters remain strongly degenerate. Its best-fit likelihood is comparable to $\Lambda$CDM, but the additional parameters are not supported by information criteria. These results show that the two thermodynamic realizations of the same nonadditive entropy have sharply different phenomenological consequences at the homogeneous-background level. In summary, the observational plausibility of entropy-based cosmology hinges critically on how the generalized horizon entropy is incorporated into the Friedmann dynamics. 
\end{abstract}


\end{frontmatter}

\section{Introduction}
\label{sec:introduction}

Cosmic acceleration was first inferred from Type-Ia supernovae and has since been corroborated by a broad set of cosmological probes \cite{SupernovaSearchTeam:1998fmf,SupernovaCosmologyProject:1998vns,SDSS:2003eyi,Peebles:2002gy}. Within general relativity, the minimal explanation is a cosmological constant, giving rise to the spatially flat $\Lambda$CDM model \cite{Peebles:2002gy,Frieman:2008sn}. Despite its impressive phenomenological success, a fundamental understanding of $\Lambda$ remains elusive, in particular its extraordinarily small value and the coincidence problem \cite{Weinberg:1988cp,Padmanabhan:2002ji,Carroll:2000fy}. In addition, the Hubble tension highlights the discrepancy between $H_0$ inferred from cosmic microwave background fits assuming flat $\Lambda$CDM and the value measured from the local Cepheid-calibrated distance ladder \cite{Planck:2018vyg,Riess:2021jrx,Verde:2019ivm}, motivating explorations of dynamical dark energy and modified gravity.

The holographic principle relates the maximum information content of a gravitating system to its boundary area \cite{tHooft:1993dmi,Susskind:1994vu}. For a black hole, this relation is encoded in the Bekenstein--Hawking entropy \cite{Hawking:1971tu,Bekenstein:1973ur,Bekenstein:1974ax,Hawking:1974rv,Hawking:1975vcx}. The connection between horizon entropy and cosmological dynamics is reinforced by the thermodynamic derivation of gravitational field equations \cite{Jacobson:1995ab} and by apparent-horizon formulations of FLRW cosmology \cite{Cai:2005ra,Akbar:2006kj,Cai:2006rs}. Related constructions include holographic equipartition \cite{Padmanabhan:2003gd,Padmanabhan:2009vy} and entropic-gravity ideas \cite{Verlinde:2010hp}.

Generalized entropy-area relations provide a natural way to explore departures from the standard Bekenstein--Hawking law. Examples include Tsallis/Tsallis--Cirto, Barrow, R\'enyi, Kaniadakis and Sharma--Mittal entropies \cite{Barboza:2014yfe,Sheykhi:2018dpn,Sheykhi:2022jqq,Komatsu:2016vof,Golanbari:2020coz,Lymperis:2021qty,Sheykhi:2023aqa,Naeem:2023tcu,Luciano:2026ufu}. Generalized entropies have also been used in holographic dark-energy constructions \cite{Tavayef:2018xwx,Saridakis:2020zol,Adhikary:2021xym,Oliveros:2022biu,Drepanou:2021jiv,Nojiri:2022aof,Upadhyay:2021atf}. These approaches are conceptually distinct, but both can generate effective dark-energy contributions through modifications of horizon physics.

Although generalized entropy frameworks can lead to theoretically interesting departures from the standard cosmological dynamics, their physical relevance depends on whether the resulting expansion history is compatible with current measurements of the late- and early-Universe background evolution. In particular, baryon acoustic oscillation, Type-Ia supernova, and cosmic microwave background distance information provide stringent tests of such scenarios and can strongly constrain the allowed deformation parameters even at the background level.

Recently, Kruglov proposed a nonadditive entropy whose equiprobable form deforms the Bekenstein--Hawking entropy as \cite{Kruglov:2025kat}
\begin{equation}
S_K=\frac{S_{\rm BH}}{1+\gamma S_{\rm BH}},
\label{eq:kruglov_intro}
\end{equation}
where $\gamma$ is a dimensionless deformation parameter and the standard entropy is recovered for $\gamma\to0$. Big Bang nucleosynthesis constraints on the cosmology associated with this entropy have recently been investigated by Phukan et al.~\cite{Phukan:2026sjh}, providing a complementary test of its early-Universe behaviour. In this work we examine two cosmological realizations of Eq.~\eqref{eq:kruglov_intro}. Model I follows Kruglov's original proposal using the Cai--Kim apparent-horizon construction and modifies the background Friedmann dynamics directly. Model II, proposed in this work, instead uses an entropy-integral prescription to construct an effective vacuum component. Although both models originate from the same entropy, their background phenomenology is very different.

Our aim is twofold. First, we formulate the models in a dimensionally consistent and numerically stable form, retaining radiation and imposing exact closure at $z=0$. Second, we confront their homogeneous-background predictions with DESI DR2 BAO and compressed CMB information, with and without DES-Y5 supernovae. The analysis is intentionally restricted to background observables; no growth or perturbation observable is used. This separation allows us to test whether the entropy-induced background expansion is viable before a full perturbation and Boltzmann treatment is developed.

The paper is organized as follows. Section~\ref{sec:entropy} summarizes the Kruglov entropy. Section~\ref{sec:models} derives the two background models and their dimensionless evolution equations. Section~\ref{sec:data} describes the numerical implementation and observational analysis. The results are presented in Sec.~\ref{sec:results}, followed by conclusions in Sec.~\ref{sec:conclusions}.

\section{Kruglov nonadditive entropy}
\label{sec:entropy}

For microscopic probabilities $\{p_i\}$, $\sum_i p_i=1$, Kruglov's entropy is \cite{Kruglov:2025kat}
\begin{equation}
S_K=-\sum_{i=1}^{W}\frac{p_i\ln p_i}{1-\gamma\ln p_i}.
\label{eq:kruglov_statistical}
\end{equation}
For equiprobable states, $p_i=1/W$, and identifying the resulting entropy with the horizon entropy, one obtains
\begin{equation}
S_K=\frac{S_{\rm BH}}{1+\gamma S_{\rm BH}},\qquad
S_{\rm BH}=\frac{A}{4G},\qquad A=4\pi r_h^2.
\label{eq:kruglov_entropy}
\end{equation}
The deformation is nonadditive for statistically independent subsystems and continuously approaches the Bekenstein--Hawking form as $\gamma\to0$.

For comparison, the Sharma--Mittal entropy $S_{\rm SM}$ can be written in term of Tsallis entropy $S_{T}$ as
\begin{equation}
S_{\rm SM}=\frac{1}{R}\left[(1+\delta S_T)^{R/\delta}-1\right],
\qquad
S_T=\frac{A}{4G}\left(\frac{A}{A_0}\right)^{\zeta-1}.
\end{equation}
The Kruglov horizon entropy in Eq.~\eqref{eq:kruglov_entropy} is recovered from this expression by setting $R=-\gamma$, $\delta=\gamma$, and $\zeta=1$.

\section{Cosmological models}
\label{sec:models}

We consider a spatially flat FLRW background and separately conserved pressureless matter and radiation,
\begin{equation}
\dot\rho_m+3H\rho_m=0,\qquad
\dot\rho_r+4H\rho_r=0.
\end{equation}
Radiation is retained because the BAO and compressed-CMB observables depend on the sound horizon. We use $c=\hbar=k_{\rm B}=1$ in the theoretical derivations.

\subsection{Model I: apparent-horizon thermodynamics}

The apparent-horizon radius of FLRW geometry is given by
\begin{equation}
\tilde r_A^2=\frac{1}{H^2+k/a^2},
\end{equation}
so that $\tilde r_A=H^{-1}$ for $k=0$. In the Cai--Kim construction, replacing the Bekenstein--Hawking entropy by Eq.~\eqref{eq:kruglov_entropy} changes the horizon entropy derivative by the factor $(1+\gamma S_{\rm BH})^{-2}$. Since $S_{\rm BH}=\pi/(GH^2)$, it is convenient to define the dimensionful entropy scale
\begin{equation}
\Gamma\equiv\frac{\gamma}{G},\qquad b\equiv\pi\Gamma=\frac{\pi\gamma}{G}.
\label{eq:Gamma_def}
\end{equation}
The modified Raychaudhuri equation is then \cite{Kruglov:2025kat}
\begin{equation}
\frac{\dot H}{(1+b/H^2)^2}=-4\pi G(\rho+p),
\label{eq:model1_raychaudhuri}
\end{equation}
where $\rho$ and $p$ denote the standard matter-radiation sector in the thermodynamic derivation. Combining Eq.~\eqref{eq:model1_raychaudhuri} with total conservation and integrating gives
\begin{equation}
H^2=\frac{8\pi G}{3}\rho+
\frac{b^2}{H^2+b}
+2b\ln\!\left(\frac{H^2+b}{b}\right)+C_{\rm int}.
\label{eq:model1_general_friedmann}
\end{equation}
In Model I we consider $C_{\rm int}=0$ \cite{Kruglov:2025kat}, so that the late-time modification is generated entirely by the entropy correction rather than by an additional cosmological-constant term. The effective dark-energy density is therefore
\begin{equation}
\rho_D=\frac{3}{8\pi G}\left[
\frac{b^2}{H^2+b}
+2b\ln\!\left(\frac{H^2+b}{b}\right)\right].
\label{eq:model1_density}
\end{equation}
Treating this effective component as conserved, $\dot\rho_D+3H(\rho_D+p_D)=0$, gives
\begin{equation}
p_D=-\rho_D-\frac{b(b+2H^2)\dot H}{4\pi G(H^2+b)^2},
\label{eq:model1_pressure}
\end{equation}
and hence
\begin{equation}
 w_D=-1-
 \frac{2(b+2H^2)\dot H}{3(H^2+b)^2}
 \left[
 \frac{b}{H^2+b}+2\ln\!\left(\frac{H^2+b}{b}\right)
 \right]^{-1}.
\label{eq:model1_w_dimensional}
\end{equation}
For $b>0$ and ordinary matter/radiation, Eq.~\eqref{eq:model1_raychaudhuri} gives $\dot H<0$, and therefore the effective component lies on the non-phantom side, $w_D>-1$.

For the numerical analysis we define
\begin{equation}
x(z)\equiv E^2(z)=\frac{H^2(z)}{H_0^2},\qquad
y\equiv\ln(1+z),\qquad B\equiv\frac{b}{H_0^2},
\end{equation}
and
\begin{equation}
M(z)=\Omega_{m0}(1+z)^3+\Omega_{r0}(1+z)^4,
\qquad
\Omega_{D0}=1-\Omega_{m0}-\Omega_{r0}.
\end{equation}
Equation~\eqref{eq:model1_general_friedmann} with $C_{\rm int}=0$ becomes
\begin{equation}
x=M(z)+g_B(x),\qquad
g_B(x)=\frac{B^2}{x+B}+2B\ln\!\left(1+\frac{x}{B}\right).
\label{eq:model1_dimensionless}
\end{equation}
The entropy parameter is not sampled independently. Instead, $B$ is determined for every cosmological sample from the exact present-day closure condition
\begin{equation}
g_B(1)=\Omega_{D0}.
\label{eq:model1_closure}
\end{equation}
The derivative
\begin{equation}
g_{B,x}=\frac{B(2x+B)}{(x+B)^2},\qquad
1-g_{B,x}=\frac{x^2}{(x+B)^2}>0
\label{eq:model1_derivative}
\end{equation}
shows that the positive implicit branch is unique. Differentiating Eq.~\eqref{eq:model1_dimensionless} with respect to $y$ gives
\begin{equation}
\frac{dx}{dy}=
\frac{3\Omega_{m0}(1+z)^3+4\Omega_{r0}(1+z)^4}
{1-g_{B,x}}.
\label{eq:model1_xprime}
\end{equation}
The deceleration parameter and effective dark-energy equation of state are
\begin{equation}
q=-1+\frac{1}{2x}\frac{dx}{dy},\qquad
w_D=-1+\frac{g_{B,x}}{3g_B(x)}\frac{dx}{dy}.
\label{eq:model1_qw}
\end{equation}

\subsection{Model II: entropy-integral dark energy}
A different construction, proposed by Manoharan \textit{et al.} \cite{Manoharan:2022qll}, derives an effective dark-energy component from the correction terms generated by a generalized horizon entropy in the thermodynamic Friedmann equations. Thus, instead of directly postulating the conventional holographic relation $\rho_\Lambda\propto S/L^4$, the construction begins with the energy density obtained from the Clausius relation or, equivalently, from the unified first law at the apparent horizon.  For a $(3+1)$-dimensional FLRW spacetime, the relation can be written as
\begin{equation}
\rho=-\frac{3}{8\pi^2}
\int\frac{dS}{\tilde r_A^4}.
\label{eq:thermodynamic_density_4d}
\end{equation}

To isolate the contribution arising specifically from a generalized entropy, the standard Bekenstein--Hawking part must be removed.The effective dark-energy density associated with the non-Bekenstein--Hawking part of the entropy is therefore defined as \cite{Manoharan:2022qll}
\begin{equation}
\rho_\Lambda=
C^2\left[
-\frac{3}{8\pi^2}
\int\frac{dS}{\tilde r_A^4}
-\frac{3}{8\pi G\tilde r_A^2}
\right].
\label{eq:model2_integral}
\end{equation}
The subtraction guarantees that, when $S=S_{\rm BH}$, the two horizon-dependent terms cancel and only the integration constant remains. Consequently, the Bekenstein--Hawking limit reproduces a cosmological-constant contribution rather than an additional term proportional to $H^2$. Applying Kruglov entropy \eqref{eq:kruglov_entropy}, a new dark energy density is obtained,
\begin{equation}
\rho_\Lambda
= \frac{C^2}{G}
\left\{
-\frac{3\Gamma}{8}
\left[
2\ln\left(
\frac{1+\pi\Gamma/H^2}{\pi\Gamma/H^2}
\right)
-\frac{1}{1+\pi\Gamma/H^2}
\right]
+\widetilde\Lambda
\right\},
\label{eq:model2_density}
\end{equation}
where $\widetilde\Lambda$ is an integration constant with dimensions of $H^2$. $\Gamma$ denotes the dimensionful combination introduced in Eq.~\eqref{eq:Gamma_def}. The limit $\Gamma\to 0$ leaves the constant contribution $C^2\widetilde\Lambda$.

Direct differentiation yields
\begin{equation}
\dot\rho_\Lambda=-\frac{3C^2\Gamma}{4GH}
\frac{2+\pi\Gamma/H^2}{(1+\pi\Gamma/H^2)^2}\dot H,
\label{eq:model2_rhodot}
\end{equation}
and, from $\dot\rho_\Lambda+3H(\rho_\Lambda+p_\Lambda)=0$,
\begin{equation}
p_\Lambda=-\rho_\Lambda+
\frac{C^2\Gamma}{4GH^2}
\frac{2+\pi\Gamma/H^2}{(1+\pi\Gamma/H^2)^2}\dot H.
\label{eq:model2_pressure}
\end{equation}
The dimensional equation of state follows directly from Eqs.~\eqref{eq:model2_density} and \eqref{eq:model2_pressure}; for inference it is simpler to use the normalized form below.

We introduce two dimensionless scale combinations,
\begin{equation}
\beta\equiv\frac{\Gamma}{\widetilde\Lambda},\qquad
\lambda\equiv\frac{\widetilde\Lambda}{H_0^2},\qquad
\frac{\Gamma}{H_0^2}=\beta\lambda.
\label{eq:model2_scales}
\end{equation}
Factoring out $\widetilde\Lambda$ defines the shape function
\begin{equation}
\Phi(x;\beta,\lambda)=1-\frac{3\beta}{8}\left[
2\ln\!\left(1+\frac{x}{\pi\beta\lambda}\right)
-\frac{x}{x+\pi\beta\lambda}\right],
\label{eq:model2_phi}
\end{equation}
with the continuous $\beta=0$ limit $\Phi=1$. Present-day closure eliminates the overall amplitude and gives
\begin{equation}
x=M(z)+\Omega_{D0}\frac{\Phi(x;\beta,\lambda)}{\Phi(1;\beta,\lambda)}.
\label{eq:model2_dimensionless}
\end{equation}
The physical present-day branch requires $\Phi(1)>0$. The analytic derivative is
\begin{equation}
\Phi_{,x}=-\frac{3\beta}{8}
\frac{2x+\pi\beta\lambda}{(x+\pi\beta\lambda)^2}.
\label{eq:model2_derivative}
\end{equation}
Writing $A\equiv\Omega_{D0}/\Phi(1)$, the background evolution becomes
\begin{equation}
\frac{dx}{dy}=\frac{3\Omega_{m0}(1+z)^3+4\Omega_{r0}(1+z)^4}
{1-A\Phi_{,x}},
\label{eq:model2_xprime}
\end{equation}
with
\begin{equation}
w_\Lambda=-1+\frac{\Phi_{,x}}{3\Phi(x)}\frac{dx}{dy},\qquad
q=-1+\frac{1}{2x}\frac{dx}{dy}.
\label{eq:model2_qw}
\end{equation}
For $\beta>0$, $\Phi_{,x}<0$. Consequently, as long as $\Phi>0$ on the regular branch, Model II lies on the phantom side, $w_\Lambda<-1$, approaching $w=-1$ continuously as $\beta\to0$. At sufficiently high redshift $\Phi$ can cross zero for part of the allowed parameter space. The ratio $w_\Lambda=p_\Lambda/\rho_\Lambda$ then diverges, but the implicit Friedmann solution can remain regular because the effective density itself passes smoothly through zero.

\section{Numerical implementation and data}
\label{sec:data}

\subsection{Background observables}

The implicit Friedmann equations are solved on a grid in $y=\ln(1+z)$. Numerically we evolve $\ln x$ using Eqs.~\eqref{eq:model1_xprime} and \eqref{eq:model2_xprime}, and then project the solution back onto the corresponding algebraic Friedmann constraint with Newton updates using the analytic derivatives. This procedure keeps $x>0$ and yields a maximum relative closure residual below $10^{-14}$ for the retained samples. The implementation is incorporated in the \texttt{CosmoDS} framework \cite{Roy:2026icy}.

We use
\begin{equation}
\Omega_{\gamma0}h^2=2.469\times10^{-5},\qquad
\Omega_{r0}h^2=\Omega_{\gamma0}h^2(1+0.2271N_{\rm eff}),
\end{equation}
with $T_{\rm CMB}=2.7255\,$K and $N_{\rm eff}=3.046$. Massive neutrinos are not included as a separate background component. The flat-universe distances are
\begin{equation}
D_C=D_M=\frac{c}{H_0}\int_0^z\frac{d\tilde z}{E(\tilde z)},\qquad
D_A=\frac{D_M}{1+z},\qquad
D_L=(1+z)D_M.
\label{eq:distances}
\end{equation}
The comoving sound horizon is
\begin{equation}
r_s(z)=\int_z^\infty\frac{c\,d\tilde z}
{\sqrt{3}\,H(\tilde z)\sqrt{1+R(\tilde z)}},\qquad
R(z)=\frac{3\Omega_{b0}}{4\Omega_{\gamma0}(1+z)}.
\label{eq:sound_horizon}
\end{equation}
We define $r_d=r_s(z_d)$, $r_*=r_s(z_*)$, and $\theta_s=r_*/D_M(z_*)$. The drag and last-scattering redshifts are evaluated with the Eisenstein--Hu and Hu--Sugiyama fitting formulae \cite{Eisenstein:1997ik,Hu:1995en}. To remain within the regime in which this background-only early-Universe treatment is meaningful, samples with $|\rho_{\rm DE}/\rho_c|_{z_*}>10^{-3}$ are rejected. A full recombination and Boltzmann treatment is left for future work.

\subsection{Observational likelihoods}

We consider the two combinations
\begin{align}
{\cal D}_1&=\text{DESI DR2 BAO}+\text{compressed CMB},\\
{\cal D}_2&=\text{DES-Y5 SNe}+\text{DESI DR2 BAO}+\text{compressed CMB}.
\label{eq:data_combinations}
\end{align}
The DESI DR2 likelihood contains 13 BAO measurements and their covariance \cite{DESI:2025zgx}. In a flat background the model predictions are
\begin{equation}
D_H=\frac{c}{H(z)},\qquad
D_V=\left[zD_M^2D_H\right]^{1/3},
\end{equation}
and the appropriate measured combinations $D_M/r_d$, $D_H/r_d$, or $D_V/r_d$ are used for each tracer bin.

The compressed CMB likelihood is a three-dimensional Gaussian in
\begin{equation}
{\bf v}=(\theta_s,\omega_b,\omega_{bc}),\qquad
\omega_b=\Omega_{b0}h^2,\qquad
\omega_{bc}=\Omega_{m0}h^2,
\label{eq:cmb_vector}
\end{equation}
using the mean vector
\begin{equation}
{\boldsymbol\mu}_{\rm CMB}=(0.01041,\,0.0222,\,0.14208)
\end{equation}
and the full covariance matrix implemented in the likelihood. For a residual vector $\Delta{\bf v}={\bf v}-{\boldsymbol\mu}_{\rm CMB}$,
\begin{equation}
\chi^2_{\rm CMB}=\Delta{\bf v}^{\rm T}C_{\rm CMB}^{-1}\Delta{\bf v}.
\end{equation}
The DES-Y5 likelihood contains 1820 SNe and uses the luminosity-distance Hubble-diagram shape with analytic marginalization over the additive magnitude intercept \cite{DES:2024jxu}. The supernovae therefore constrain relative distances without introducing a sampled absolute-magnitude parameter. A compilation of 31 direct cosmic-chronometer measurements is displayed only as a visual comparison in the background-history figures and does not enter either $\mathcal D_1$ or $\mathcal D_2$ .

The posterior is
\begin{equation}
P(\vartheta|D,M)\propto{\cal L}(D|\vartheta,M)P(\vartheta|M),\qquad
\ln{\cal L}_{\rm tot}=\sum_i\ln{\cal L}_i.
\end{equation}
We impose uniform priors
\begin{equation}
50\le H_0\le90,\quad
0.03\le\Omega_{b0}\le0.07,\quad
0.05\le\Omega_{m0}\le0.60,
\label{eq:common_priors}
\end{equation}
with $H_0$ in km\,s$^{-1}$\,Mpc$^{-1}$ and $\Omega_{b0}<\Omega_{m0}$. Model I has no additional sampled parameter because $B$ is fixed by Eq.~\eqref{eq:model1_closure}. For Model II we sample
\begin{equation}
0\le\beta\le5,\qquad0.05\le\lambda\le20,
\label{eq:model2_priors}
\end{equation}
with the additional physical requirement $\Phi(1)>0$. Points for which the implicit branch loses monotonicity, $1-A\Phi_{,x}\le0$, are rejected. Thus $\Lambda$CDM and Model I each have three sampled cosmological parameters, while Model II has five. Sampling is performed with the Cobaya \cite{Torrado:2020dgo} Metropolis algorithm using six MPI chains and the convergence criterion $R-1<0.02$.

\section{Results}
\label{sec:results}

Table~\ref{tab:background_constraints} summarizes the parameters common to all models. For $\mathcal D_1$, flat $\Lambda$CDM gives $H_0=66.710^{+0.283}_{-0.250}$ km\,s$^{-1}$\,Mpc$^{-1}$ and $\Omega_{m0}=0.32564^{+0.00368}_{-0.00391}$. Model I is clearly displaced, with $H_0=62.127^{+0.244}_{-0.245}$ km\,s$^{-1}$\,Mpc$^{-1}$ and $\Omega_{m0}=0.36516^{+0.00446}_{-0.00459}$. Adding DES-Y5 shifts Model I only modestly to $H_0=62.391^{+0.219}_{-0.239}$ km\,s$^{-1}$\,Mpc$^{-1}$. This is a direct consequence of the exact closure condition: once $B$ is fixed, the standard cosmological parameters must compensate the altered distance-redshift relation.

Model II behaves very differently. Its posterior in $(H_0,\Omega_{b0},\Omega_{m0})$ largely overlaps that of $\Lambda$CDM. For $\mathcal D_2$ we find $H_0=66.816^{+0.307}_{-0.291}$ km\,s$^{-1}$\,Mpc$^{-1}$ and $\Omega_{m0}=0.32647^{+0.00395}_{-0.00409}$. The present equation of state remains only slightly phantom, $w_{\rm DE,0}=-1.00263^{+0.00053}_{-0.00145}$. The corresponding Model-I value, $w_{\rm DE,0}=-0.81370^{+0.00125}_{-0.00118}$, is substantially different from $-1$ and is fully consistent with the analytical sign implied by Eq.~\eqref{eq:model1_w_dimensional}.

\begin{table*}[t]
\centering
\scriptsize
\renewcommand{\arraystretch}{1.18}
\caption{Marginalized background constraints. Entries are posterior medians with central 68\% intervals. $r_d$ is in Mpc and $H_0$ is in km\,s$^{-1}$\,Mpc$^{-1}$.}
\label{tab:background_constraints}
\resizebox{\textwidth}{!}{%
\begin{tabular}{llccccccc}
\toprule
Data combination & Model & $H_0$ & $\Omega_{b0}$ & $\Omega_{m0}$ & $\Omega_{\rm DE,0}$ & $w_{\rm DE,0}$ & $q_0$ & $r_d$\\
\midrule
\multirow{3}{*}{\shortstack{DESI DR2 BAO\\+ compressed CMB}}
 & $\Lambda$CDM & $66.710^{+0.283}_{-0.250}$ & $0.04944^{+0.00033}_{-0.00034}$ & $0.32564^{+0.00368}_{-0.00391}$ & $0.67426^{+0.00391}_{-0.00368}$ & $-1$ (fixed) & $-0.51135^{+0.00552}_{-0.00586}$ & $150.417^{+0.188}_{-0.196}$\\
 & Model I & $62.127^{+0.244}_{-0.245}$ & $0.05785^{+0.00034}_{-0.00035}$ & $0.36516^{+0.00446}_{-0.00459}$ & $0.63473^{+0.00459}_{-0.00446}$ & $-0.81228^{+0.00130}_{-0.00136}$ & $-0.27332^{+0.00666}_{-0.00690}$ & $151.222^{+0.205}_{-0.198}$\\
 & Model II & $66.942^{+0.315}_{-0.304}$ & $0.04900^{+0.00037}_{-0.00040}$ & $0.32466^{+0.00429}_{-0.00406}$ & $0.67524^{+0.00407}_{-0.00429}$ & $-1.00270^{+0.00061}_{-0.00169}$ & $-0.51584^{+0.00652}_{-0.00688}$ & $150.291^{+0.198}_{-0.202}$\\
\midrule
\multirow{3}{*}{\shortstack{DES-Y5 + DESI DR2 BAO\\+ compressed CMB}}
 & $\Lambda$CDM & $66.617^{+0.295}_{-0.288}$ & $0.04951^{+0.00033}_{-0.00032}$ & $0.32697^{+0.00413}_{-0.00414}$ & $0.67293^{+0.00414}_{-0.00413}$ & $-1$ (fixed) & $-0.50936^{+0.00620}_{-0.00622}$ & $150.394^{+0.172}_{-0.185}$\\
 & Model I & $62.391^{+0.219}_{-0.239}$ & $0.05755^{+0.00033}_{-0.00032}$ & $0.36036^{+0.00422}_{-0.00392}$ & $0.63953^{+0.00392}_{-0.00422}$ & $-0.81370^{+0.00125}_{-0.00118}$ & $-0.28052^{+0.00634}_{-0.00593}$ & $151.346^{+0.190}_{-0.201}$\\
 & Model II & $66.816^{+0.307}_{-0.291}$ & $0.04912^{+0.00036}_{-0.00038}$ & $0.32647^{+0.00395}_{-0.00409}$ & $0.67344^{+0.00409}_{-0.00395}$ & $-1.00263^{+0.00053}_{-0.00145}$ & $-0.51322^{+0.00618}_{-0.00640}$ & $150.254^{+0.200}_{-0.190}$\\
\bottomrule
\end{tabular}}
\end{table*}

The model-specific constraints are shown in Table~\ref{tab:model_specific}. In Model I, $B$ is derived from closure and is tightly localized near $0.15$. The dimensional quantity $b=BH_0^2$ is therefore also narrow, but $B$ and $b$ are not independent degrees of freedom. In Model II the posterior in $(\beta,\lambda)$ is broad and strongly correlated. The one-dimensional intervals are visibly affected by the prior boundaries, and therefore should not be interpreted as independent detections of either entropy-sector scale.

\begin{table}[t]
\centering
\scriptsize
\renewcommand{\arraystretch}{1.15}
\caption{Constraints on model-specific parameters. $B$ is dimensionless and $b$ is in km$^2$\,s$^{-2}$\,Mpc$^{-2}$.}
\label{tab:model_specific}
\begin{tabular}{llcccc}
\toprule
Data & Model & $B$ & $b$ & $\beta$ & $\lambda$\\
\midrule
\multirow{3}{*}{$\mathcal D_1$}
& $\Lambda$CDM & -- & -- & -- & --\\
& Model I & $0.15159^{+0.00179}_{-0.00173}$ & $584.997^{+11.499}_{-11.058}$ & -- & --\\
& Model II & -- & -- & $1.408^{+2.283}_{-1.197}$ & $14.139^{+4.213}_{-5.811}$\\
\midrule
\multirow{3}{*}{$\mathcal D_2$}
& $\Lambda$CDM & -- & -- & -- & --\\
& Model I & $0.15347^{+0.00154}_{-0.00165}$ & $597.376^{+10.381}_{-10.873}$ & -- & --\\
& Model II & -- & -- & $1.423^{+2.258}_{-1.217}$ & $14.590^{+3.825}_{-5.315}$\\
\bottomrule
\end{tabular}
\end{table}

\begin{figure*}[t]
\centering
\includegraphics[width=0.92\textwidth]{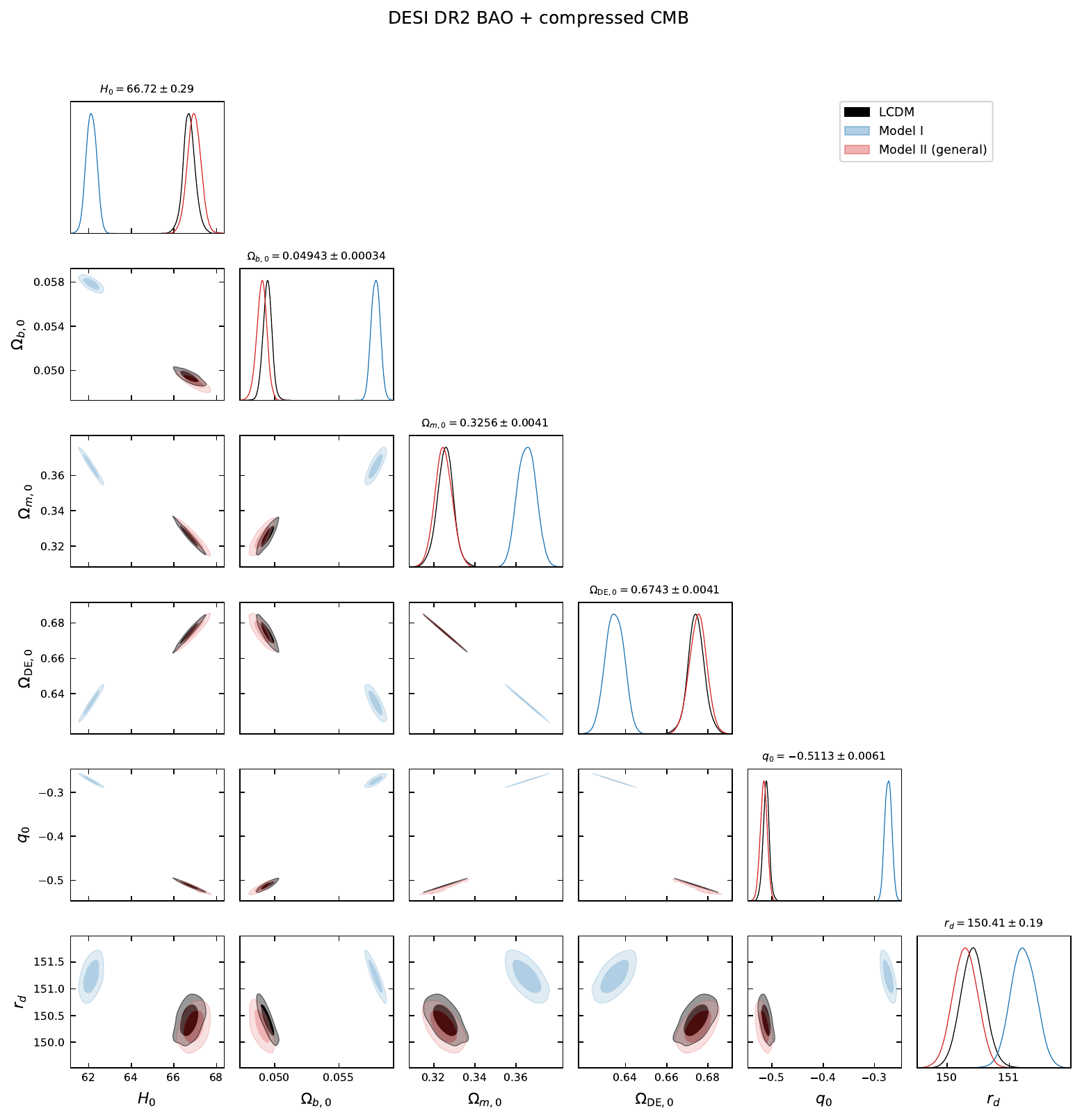}
\caption{Marginalized posterior distributions for the parameters common to all models, inferred from $\mathcal D_1$ (DESI DR2 BAO + compressed CMB). The diagonal panels show the one-dimensional marginalized posteriors for $(H_0,\Omega_{b0},\Omega_{m0},\Omega_{\rm DE,0},q_0,r_d)$, while the off-diagonal panels show the corresponding two-dimensional joint constraints (68\% and 95\% credible regions). Black, blue, and red correspond to $\Lambda$CDM, Model I, and Model II, respectively.}
\label{fig:triangle_d1}
\end{figure*}

\begin{figure*}[t]
\centering
\includegraphics[width=0.92\textwidth]{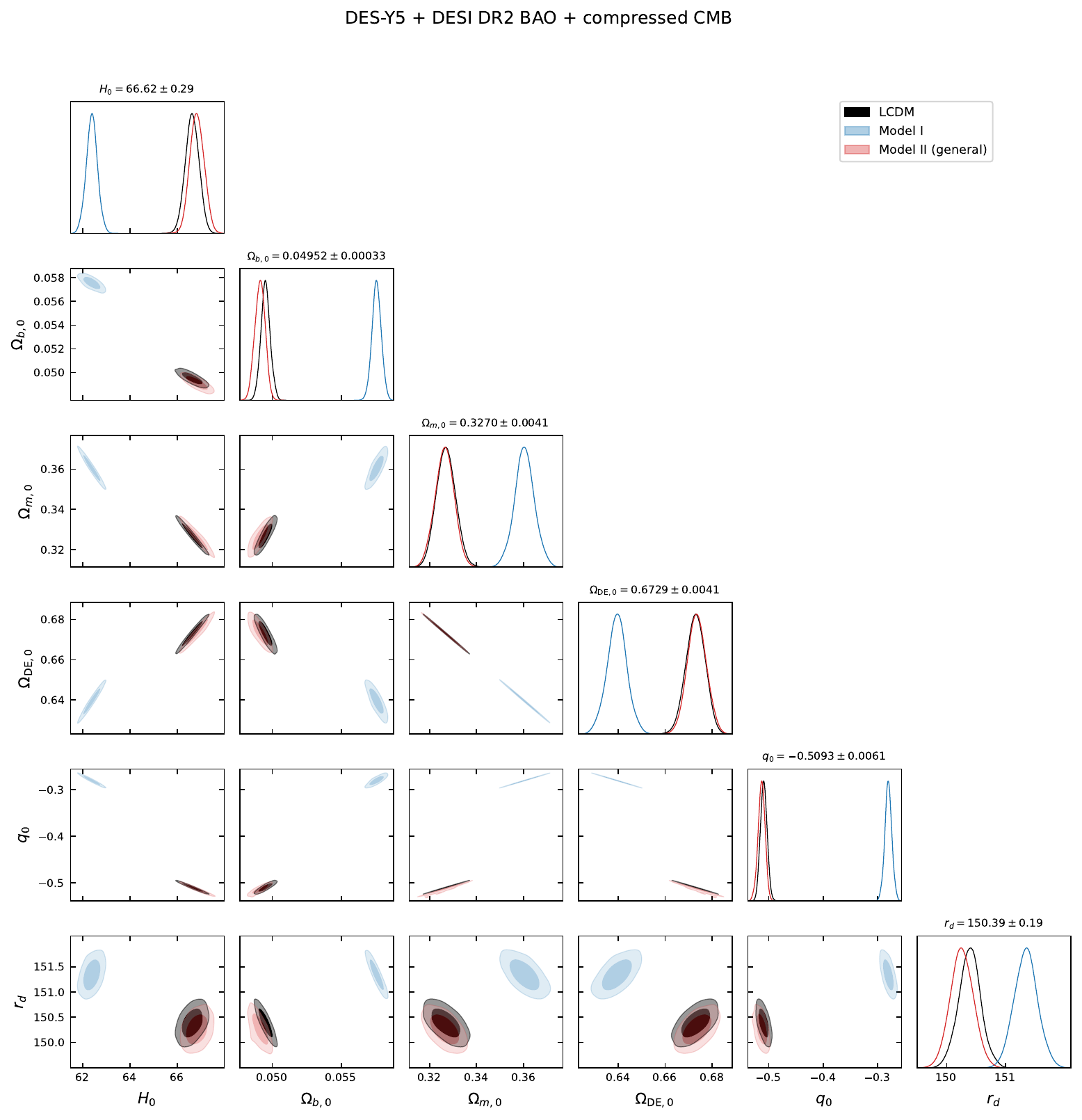}
\caption{Marginalized posterior distributions for the parameters common to all models, inferred from $\mathcal D_2$ (DES-Y5 + DESI DR2 BAO + compressed CMB). The diagonal panels show the one-dimensional marginalized posteriors for $(H_0,\Omega_{b0},\Omega_{m0},\Omega_{\rm DE,0},q_0,r_d)$, while the off-diagonal panels show the corresponding two-dimensional joint constraints (68\% and 95\% credible regions). Black, blue, and red correspond to $\Lambda$CDM, Model I, and Model II, respectively.}
\label{fig:triangle_d2}
\end{figure*}

Figures~\ref{fig:triangle_d1} and \ref{fig:triangle_d2} show the common-parameter posteriors. The separation of Model I from $\Lambda$CDM is driven mainly by the $H_0$--$\Omega_{m0}$ relation, while Model II largely follows the $\Lambda$CDM contours. The sound horizon changes only modestly, with Model I preferring slightly larger $r_d$ despite its much lower $H_0$.

\begin{figure*}[t]
\centering
\includegraphics[width=0.92\textwidth]{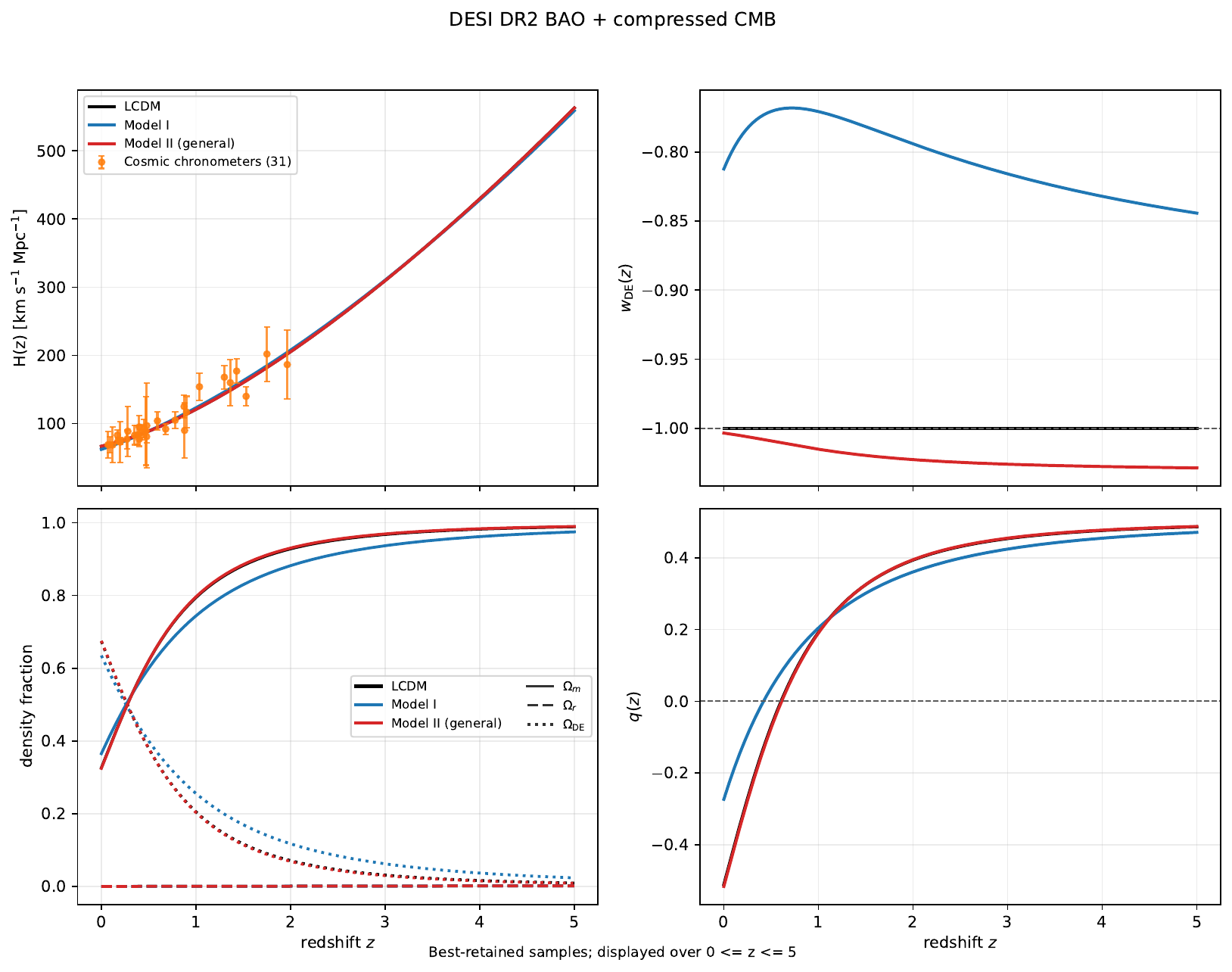}
\caption{Reconstructed background evolution for the best-fit (maximum-likelihood) samples obtained from $\mathcal D_1$ (DESI DR2 BAO + compressed CMB). Top left: expansion history $H(z)$ (cosmic-chronometer data are overplotted for visual comparison only and are not included in $\mathcal D_1$). Top right: effective dark-energy equation of state $w_{\rm DE}(z)$ (horizontal line: $w=-1$). Bottom left: reconstructed density fractions $\Omega_i(z)$ for matter, radiation, and the effective dark-energy component. Bottom right: deceleration parameter $q(z)$ (horizontal line: $q=0$).}
\label{fig:background_d1}
\end{figure*}

\begin{figure*}[t]
\centering
\includegraphics[width=0.92\textwidth]{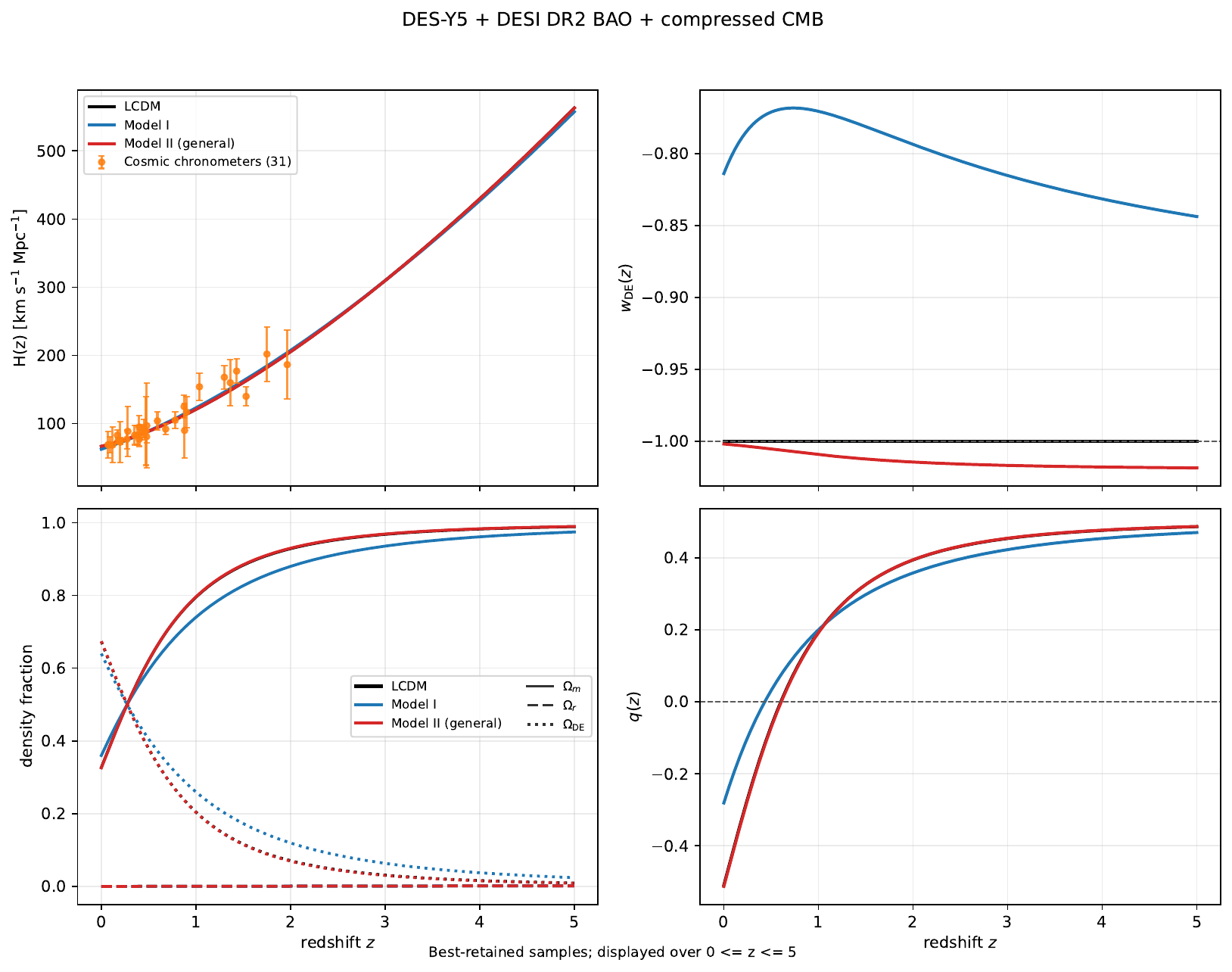}
\caption{Reconstructed background evolution for the best-fit (maximum-likelihood) samples obtained from $\mathcal D_2$ (DES-Y5 + DESI DR2 BAO + compressed CMB). Top left: expansion history $H(z)$ (cosmic-chronometer data are overplotted for visual comparison only and are not included in $\mathcal D_2$). Top right: effective dark-energy equation of state $w_{\rm DE}(z)$ (horizontal line: $w=-1$). Bottom left: reconstructed density fractions $\Omega_i(z)$ for matter, radiation, and the effective dark-energy component. Bottom right: deceleration parameter $q(z)$ (horizontal line: $q=0$).}
\label{fig:background_d2}
\end{figure*}

The reconstructed histories are shown in Figs.~\ref{fig:background_d1} and \ref{fig:background_d2}. Model I remains non-phantom over the displayed late-time interval and predicts a considerably less negative present deceleration parameter than $\Lambda$CDM. Model II remains very close to $\Lambda$CDM at low redshift. The effective density fractions shown in the figures are reconstructed from the implicit Friedmann equation, rather than obtained by evolving a separate fundamental dark-energy fluid.

\begin{figure*}[t]
\centering
\begin{subfigure}[b]{0.48\textwidth}
\centering
\includegraphics[width=\textwidth]{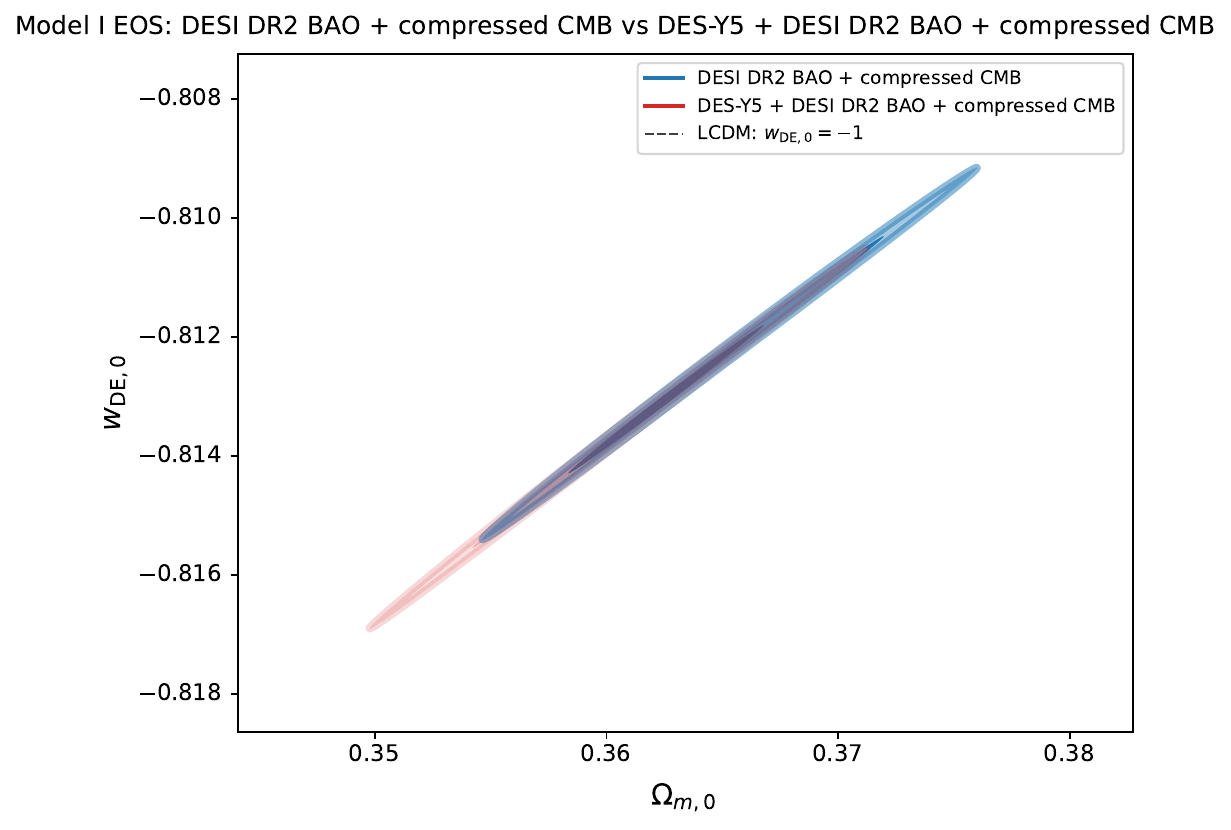}
\caption{Model I.}
\end{subfigure}\hfill
\begin{subfigure}[b]{0.48\textwidth}
\centering
\includegraphics[width=\textwidth]{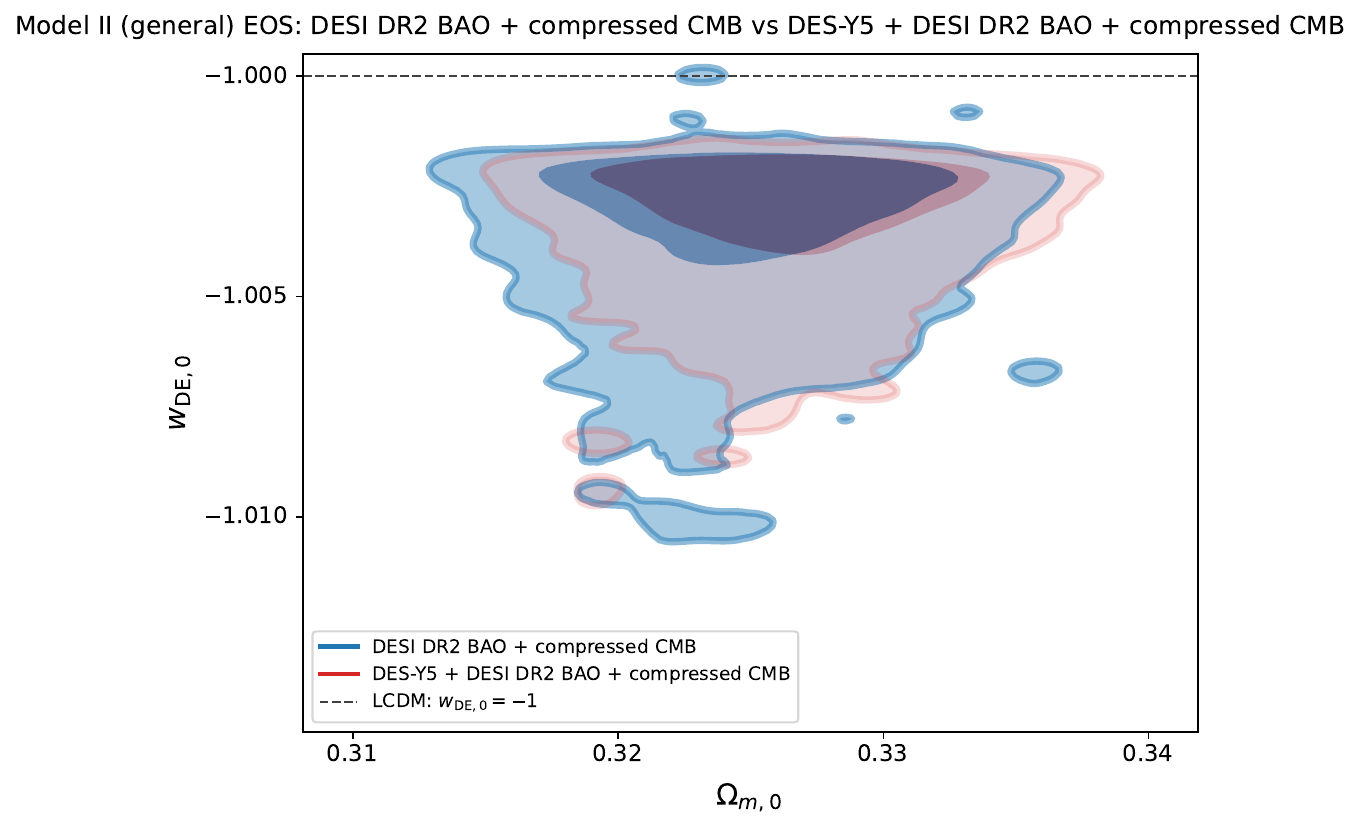}
\caption{Model II.}
\end{subfigure}
\caption{Posterior constraints in the $(\Omega_{m0},w_{\rm DE,0})$ plane. Left and right panels correspond to Model I and Model II, respectively. Blue and red contours show the 68\% and 95\% credible regions obtained from $\mathcal D_1$ and $\mathcal D_2$, respectively. The dashed horizontal line marks the $\Lambda$CDM boundary $w_{\rm DE,0}=-1$.}
\label{fig:eos_contours}
\end{figure*}

\begin{figure*}[t]
\centering
\begin{subfigure}[b]{0.48\textwidth}
\centering
\includegraphics[width=\textwidth]{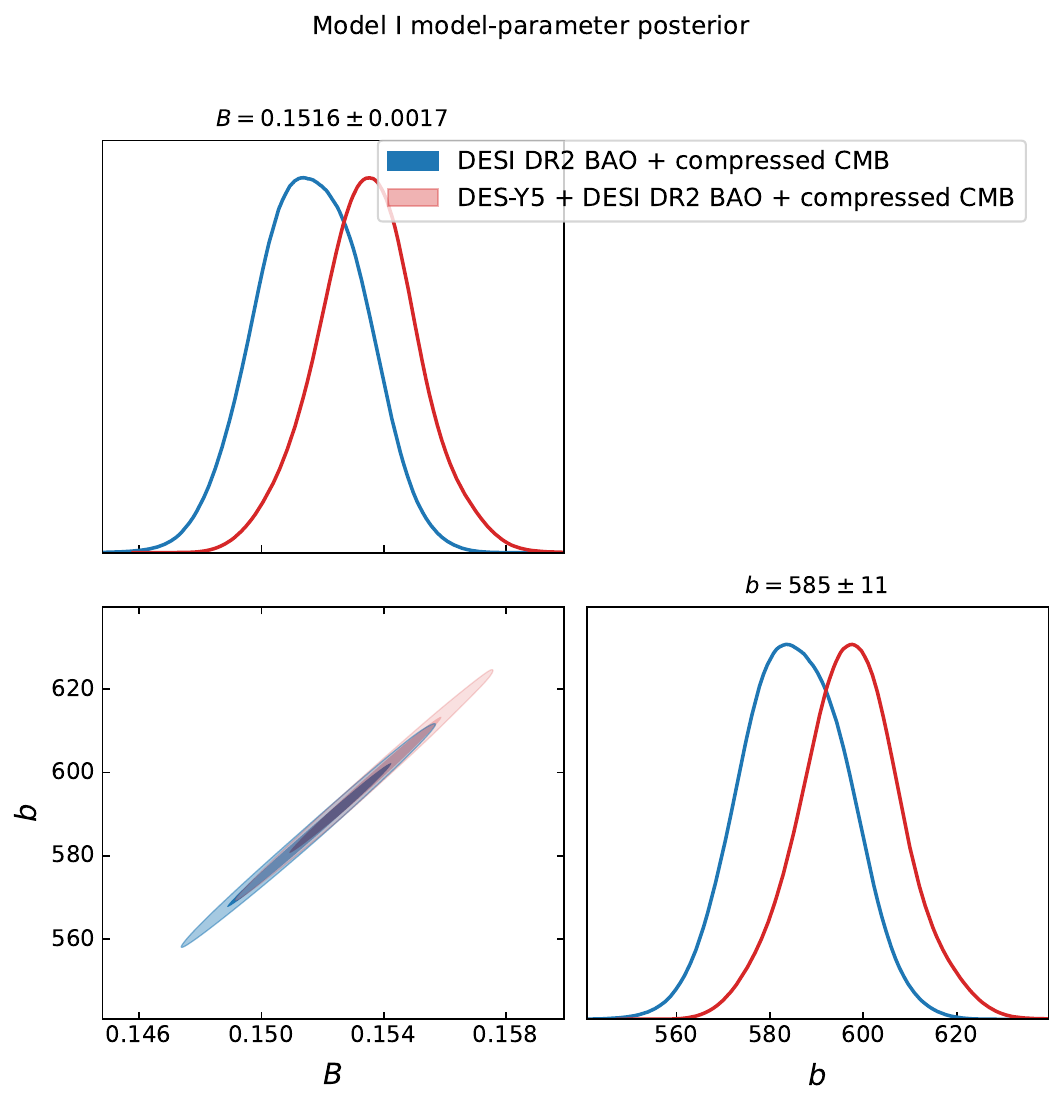}
\caption{Model I: posterior for the derived parameters $B$ and $b\equiv BH_0^2$.}
\end{subfigure}\hfill
\begin{subfigure}[b]{0.48\textwidth}
\centering
\includegraphics[width=\textwidth]{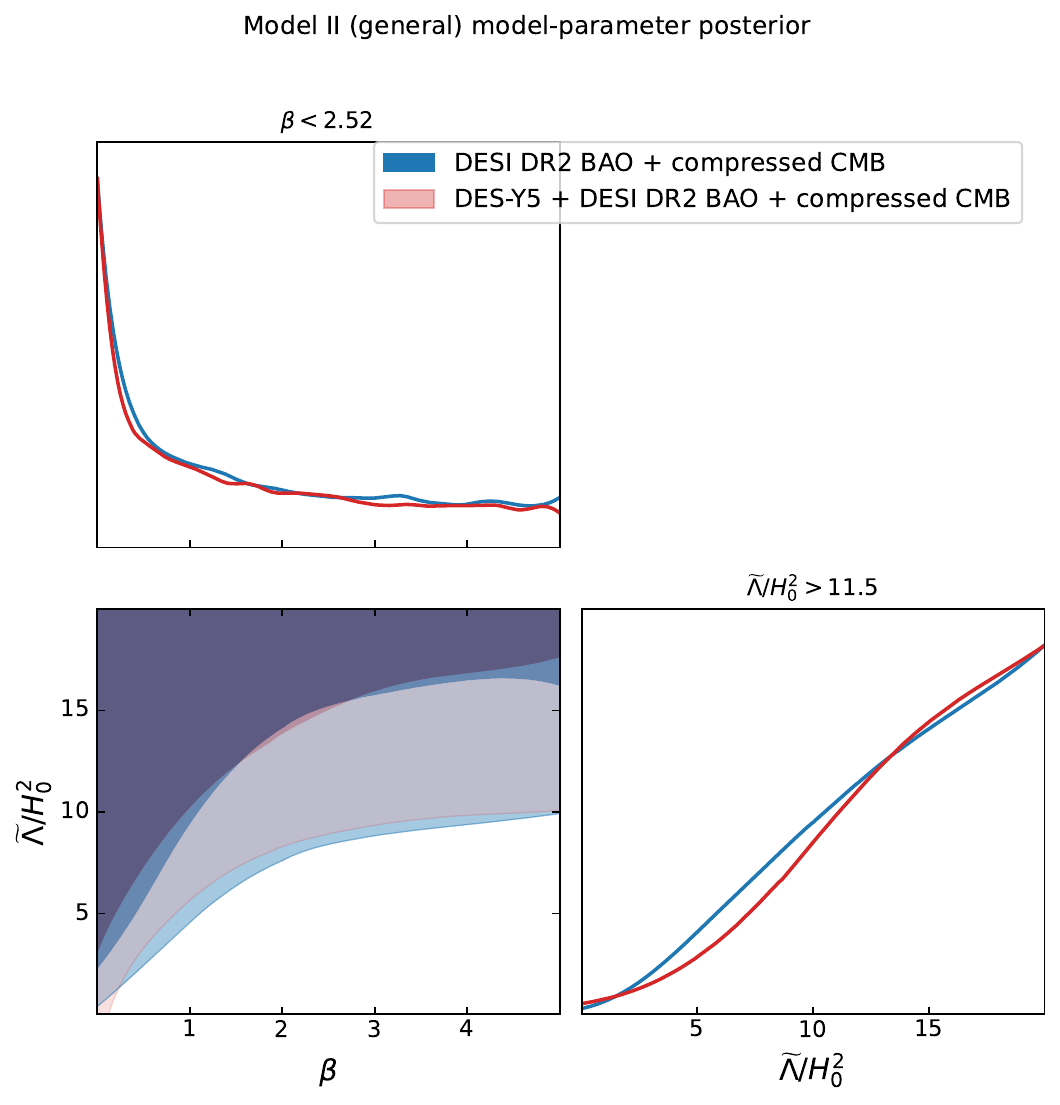}
\caption{Model II: posterior for the sampled parameters $(\beta,\lambda)$.}
\end{subfigure}
\caption{Model-specific parameter constraints. Left panel (Model I): one-dimensional marginalized posteriors for $B$ and $b\equiv BH_0^2$ (top row and bottom right) and their joint two-dimensional constraint (bottom left). Right panel (Model II): one-dimensional marginalized posteriors for $\beta$ and $\lambda$ (top row and bottom right) and their joint two-dimensional constraint (bottom left). Blue and red show results from $\mathcal D_1$ and $\mathcal D_2$, respectively; shaded regions denote the 68\% and 95\% credible intervals/regions.}
\label{fig:model_params}
\end{figure*}

Figure~\ref{fig:eos_contours} compares the two data combinations in the $(\Omega_{m0},w_{\rm DE,0})$ plane. Model I occupies a narrow region entirely above $w=-1$. Model II approaches the $\Lambda$CDM boundary from the phantom side; within the physical positive-density branch its analytical evolution does not cross to $w>-1$. Figure~\ref{fig:model_params} displays the corresponding model-specific posterior geometry.

\subsection{Model comparison}

We use
\begin{equation}
\mathrm{AIC}=-2\ln{\cal L}_{\max}+2k,\qquad
\mathrm{BIC}=-2\ln{\cal L}_{\max}+k\ln N,
\end{equation}
where $k$ is the number of sampled cosmological parameters. The analytically marginalized supernova intercept is common to all models and therefore does not affect the quoted differences in AIC or BIC. We use $N=16$ for $\mathcal D_1$ and $N=1836$ for $\mathcal D_2$. The smallest $\chi^2$ retained in the converged chains is used as the likelihood diagnostic.

\begin{table}[t]
\centering
\scriptsize
\setlength{\tabcolsep}{4pt}
\renewcommand{\arraystretch}{1.05}
\caption{Information-criterion comparison relative to $\Lambda$CDM for the same data combination.}
\label{tab:model_comparison}
\begin{tabular}{@{}llccc@{}}
\toprule
Data & Model & $\chi^2_{\min}$ & $\Delta$AIC & $\Delta$BIC\\
\midrule
\multirow{3}{*}{$\mathcal D_1$}
& $\Lambda$CDM & 37.531 & 0 & 0\\
& Model I & 76.795 & 39.264 & 39.264\\
& Model II & 37.490 & 3.959 & 5.504\\
\midrule
\multirow{3}{*}{$\mathcal D_2$}
& $\Lambda$CDM & 1679.973 & 0 & 0\\
& Model I & 1741.639 & 61.666 & 61.666\\
& Model II & 1680.414 & 4.441 & 15.471\\
\bottomrule
\end{tabular}
\end{table}
The distinction between the two entropy constructions is evident in Table~\ref{tab:model_comparison}. Model I has the same number of sampled parameters as $\Lambda$CDM but worsens the best-fit $\chi^2$ by $39.264$ for $\mathcal D_1$ and $61.666$ for $\mathcal D_2$. It is therefore strongly disfavoured by the background distances. Model II, in contrast, attains essentially the same best-fit likelihood as $\Lambda$CDM for $\mathcal D_1$ and only a slightly worse value for $\mathcal D_2$. Its information-criterion penalty comes primarily from the two additional entropy-sector parameters, which are only weakly constrained. The data consequently provide no evidence that the extra Model-II freedom is required.

\subsection{Kinematical diagnostics and low-redshift characterization}

Beyond the background expansion rate itself, it is useful to characterize the
two entropy-induced cosmologies through quantities that directly describe the
late-time kinematics. These diagnostics help clarify why the two realizations
of the same Kruglov entropy lead to markedly different observational behaviour.

For both models, the background evolution can be written in the implicit form
\begin{equation}
x=M(z)+{\cal D}(x),
\end{equation}
where $x=E^2=H^2/H_0^2$ and
\begin{equation}
M(z)=\Omega_{m0}(1+z)^3+\Omega_{r0}(1+z)^4.
\end{equation}
For Model I, ${\cal D}(x)=g_B(x)$, while for Model II
\begin{equation}
{\cal D}(x)=A\Phi(x), \qquad
A=\frac{\Omega_{D0}}{\Phi(1)}.
\end{equation}

Defining
\begin{equation}
{\cal Q}(x)=1-{\cal D}_{,x},
\end{equation}
one obtains
\begin{equation}
\frac{dx}{dy}
=
\frac{M_1}{\mathcal Q},
\qquad
M_1=
3\Omega_{m0}(1+z)^3
+
4\Omega_{r0}(1+z)^4,
\end{equation}
where $y=\ln(1+z)$. Differentiating once more gives
\begin{equation}
\frac{d^2x}{dy^2}
=
\frac{M_2}{\mathcal Q}
+
\frac{{\cal D}_{,xx}M_1^2}{\mathcal Q^3},
\end{equation}
with
\begin{equation}
M_2=
9\Omega_{m0}(1+z)^3
+
16\Omega_{r0}(1+z)^4.
\end{equation}

The deceleration parameter is
\begin{equation}
q(z)
=
-1+\frac{1}{2x}\frac{dx}{dy},
\end{equation}
and the transition from decelerated to accelerated expansion occurs at the
redshift $z_t$ satisfying
\begin{equation}
q(z_t)=0.
\end{equation}

A useful higher-order diagnostic is the jerk parameter,
\begin{equation}
j(z)
=
\frac{\dddot a}{aH^3},
\end{equation}
which can be expressed as
\begin{equation}
j(z)
=
1-\frac{3}{2x}\frac{dx}{dy}
+
\frac{1}{2x}\frac{d^2x}{dy^2}.
\end{equation}
In the late-time flat $\Lambda$CDM limit, neglecting the very small radiation
contribution, one has $j=1$. Deviations from unity therefore provide a simple
measure of departures from the standard expansion history.

It is also useful to characterize the effective dark-energy equation of state
around the present epoch through a local CPL form,
\begin{equation}
w(a)\simeq w_0+w_a(1-a).
\end{equation}
Since $y=\ln(1+z)$, one has at the present epoch
\begin{equation}
w_a
=
\left.\frac{dw}{dz}\right|_{z=0}
=
\left.\frac{dw}{dy}\right|_{y=0}.
\end{equation}
For an effective dark-energy density ${\cal D}(x)$,
\begin{equation}
w_{\rm DE}
=
-1+
\frac{{\cal D}_{,x}}{3{\cal D}}
\frac{dx}{dy},
\end{equation}
and therefore
\begin{equation}
w_a=
\left[
\frac{
{\cal D}_{,xx}(x')^2+
{\cal D}_{,x}x''
}
{3{\cal D}}
-
\frac{
{\cal D}_{,x}^{\,2}(x')^2
}
{3{\cal D}^{\,2}}
\right]_{z=0},
\end{equation}
where primes denote derivatives with respect to $y$.

For Model I, ${\cal D} = g_{B}(x)$
with
\begin{equation}
g_{B,xx}
=
-\frac{2Bx}{(x+B)^3}.
\end{equation}
Using the present-day closure condition $g_B(1)=\Omega_{D0}$, one obtains
\begin{equation}
w_{D0}
=
-1+
\frac{B(B+2)}
{3\Omega_{D0}}
\left(3\Omega_{m0}+4\Omega_{r0}\right).
\end{equation}
Similarly,
\begin{equation}
q_0
=
-1+
\frac{1}{2}(1+B)^2
\left(3\Omega_{m0}+4\Omega_{r0}\right).
\end{equation}
Thus, once the present matter density fixes $B$ through the closure relation,
the present equation of state is no longer an independently adjustable
quantity. This restricted freedom is an important reason why Model I is unable
to closely reproduce the $\Lambda$CDM background expansion preferred by the
BAO and CMB distance information.

For Model II,
\begin{equation}
\Phi_{,xx}
=
\frac{3\beta x}
{4\left(x+\pi\beta\lambda\right)^3}.
\end{equation}
Since $\Phi_{,x}<0$ for $\beta>0$, the positive-density branch
$\Phi>0$ satisfies
\begin{equation}
w_\Lambda<-1.
\end{equation}
Hence Model II approaches the $\Lambda$CDM value $w=-1$ from the phantom side
as $\beta\rightarrow0$. The effective density may nevertheless cross zero at
higher redshift, at which point $w_\Lambda$ becomes formally singular although
the expansion history itself remains regular.
\section{Conclusions}
\label{sec:conclusions}

We have studied two background cosmologies generated from Kruglov's nonadditive entropy at the apparent horizon. Although the models share the same entropy functional, the two thermodynamic constructions lead to very different cosmological behavior.

In Model I, the Cai--Kim construction produces an implicit modification of the Friedmann equation. After setting the independent integration constant to zero, the entropy scale $B=b/H_0^2$ is fixed uniquely by present-day closure rather than sampled as an additional parameter. The resulting effective component is necessarily non-phantom on the physical $b>0$ branch. The observational consequences are substantial: DESI DR2 BAO and compressed CMB information drive the model toward $H_0\simeq62\,\mathrm{km\,s^{-1}\,Mpc^{-1}}$ and a larger matter fraction than in $\Lambda$CDM. Adding DES-Y5 supernovae does not remove this displacement. The large degradation in $\chi^2$ shows that this realization is strongly disfavoured by the background data considered here.

Model II, constructed from the entropy integral, is much closer to $\Lambda$CDM. Its normalized background equation depends on the two scale combinations $(\beta,\lambda)$ and contains $\Lambda$CDM as the continuous $\beta=0$ limit. The standard parameters remain almost unchanged relative to $\Lambda$CDM, while the effective equation of state is only mildly phantom today. The two additional parameters are, however, strongly degenerate and prior-sensitive. The best-fit likelihood therefore does not provide evidence for a departure from $\Lambda$CDM, and the extra freedom is penalized by AIC and especially by BIC for the supernova-inclusive combination.

The present analysis is deliberately limited to the homogeneous background. The recombination and drag redshifts are evaluated with fitting formulae and the CMB information is compressed to background quantities. A definitive test of Kruglov-entropy cosmology will require a consistent perturbation theory, recombination history, and full Boltzmann calculation. Nevertheless, the background analysis already establishes an important result: the observational viability of entropy-based cosmology depends crucially on how the generalized horizon entropy is translated into the Friedmann dynamics. The entropy-integral realization can closely mimic $\Lambda$CDM, whereas the direct apparent-horizon modification considered here is strongly constrained by current distance measurements.

\section*{Acknowledgments}
BG is supported by a TRF Basic Research Grant No. BRG6080003 (TRF Advanced Research Scholar) of the Thailand Research Fund and the Royal Society-Newton Advanced Fellowship (NAF-R2-180874). CM is supported by a Royal Golden Jubilee-ASEAN Ph.D. scholarship under contract no. NRCT5-RGJ63009.

\bibliographystyle{elsarticle-num}
\bibliography{qeos}

\end{document}